\documentclass[letterpaper,twocolumn,10pt]{article}
\usepackage{subcaption}
\usepackage[table]{xcolor}
\usepackage{tikz}
\usepackage{amsmath}
\usepackage{amssymb}
\usepackage{booktabs}
\usepackage{multirow}
\usepackage{multicol}
\usepackage{filecontents}
\usepackage{mdframed}
\usepackage{url}
\usepackage{usenix,epsfig,endnotes}

\begin{document}

\title{\Large \bf Benchmarking Fake Voice Detection in the \\ Fake Voice Generation Arms Race}
\author{
  \centering
  \begin{tabular}{c}
    Xutao Mao \qquad Ke Li \qquad Cameron Baird \qquad Ezra Xuanru Tao \qquad Dan Lin\thanks{Corresponding author: dan.lin@vanderbilt.edu} \\
    \textit{Vanderbilt University}
  \end{tabular}
}

\maketitle
\begin{abstract}

The rapid advancement of fake voice generation technology has ignited a race with detection systems, creating an urgent need to secure the audio ecosystem. However, existing benchmarks suffer from a critical limitation: they typically aggregate diverse fake voice samples into a single dataset for evaluation. This practice masks method-specific artifacts and obscures the varying performance of detectors against different generation paradigms, preventing a nuanced understanding of their true vulnerabilities. To address this gap, we introduce the first ecosystem-level benchmark that systematically evaluates the interplay between 17 state-of-the-art fake voice generators and 8 leading detectors through a novel one-to-one evaluation protocol. This fine-grained analysis exposes previously hidden vulnerabilities and sensitivities that are missed by traditional aggregated testing. We also propose unified scoring systems to quantify both the evasiveness of generators and the robustness of detectors, enabling fair and direct comparisons. Our extensive cross-domain evaluation reveals that modern generators, particularly those based on neural audio codecs and flow matching, consistently evade top-tier detectors. We found that no single detector is universally robust; their effectiveness varies dramatically depending on the generator's architecture, highlighting a significant generalization gap in current defenses. This work provides a more realistic assessment of the threat landscape and offers actionable insights for building the next generation of detection systems.

\end{abstract}

\section{Introduction}

The evolution of fake voice generation has transformed digital communication, with applications from Text-to-Speech (TTS) and voice conversion to audio reconstruction achieving unprecedented naturalness and quality \cite{ge2021raw}. Early techniques based on parametric \cite{7472730,qian2014training,tokuda2000speech,tokuda2002hmm} or concatenative \cite{article,khan2016concatenative,tabet2011speech} methods have given way to deep neural architectures. Models like WaveNet \cite{2016arXiv160903499V} and Tacotron \cite{2017arXiv170310135W}, and more recently, vocoders based on generative adversarial networks (GANs) \cite{1406.2661} and diffusion models \cite{2023arXiv230313336Z,2020arXiv201005646K,2019arXiv191006711K}, have pushed the boundaries, achieving near-human levels of prosody and intelligibility.

This rapid advancement enables a wide range of beneficial services, from virtual assistants to content creation \cite{stan2021generating,mukhamadiyev2023voice}. However, as synthetic voices approach human parity, they also pose new and serious threats across various sectors. Beyond challenging the integrity of audio evidence in forensics, high-fidelity voice cloning is actively used for sheer deception. Malicious actors have targeted political figures \cite{bbc_biden_2024} and facilitated large-scale financial fraud by impersonating executives \cite{trendmicro2019deepfake}. These incidents highlight critical vulnerabilities in systems that rely on voice for authentication and verification.

Consequently, an intense race has emerged between advanced fake voice generation techniques and the automated detection systems designed to counter them \cite{Cheng2023AnalysisOS}. These detectors are critical Counter-Measures (CM) for technologies like Automatic Speaker Verification (ASV), which are vulnerable to spoofing attacks \cite{gupta2024vulnerability,yu2023antifake}. While prior studies have evaluated detection systems \cite{2021arXiv211203099A,2023arXiv230313336Z,2021arXiv210615561T,2024arXiv240915180P,2024arXiv240413914L}, they have a significant limitation: they conventionally aggregate fake voice samples into a single, homogeneous dataset for training and evaluation \cite{2024arXiv240808739W,2210.02437,2023arXiv230513774Y,8906599,2022arXiv220712308M,2019arXiv191101601W,2024arXiv240917285J,2023arXiv230906014W,2022arXiv221010570W}. This practice ignores critical nuances, preventing the identification of unique acoustic artifacts inherent to each generation paradigm, such as those from different voice conversion or synthesis models, and masking variations in detector performance across generator types. Recent work has shown that acoustic models and vocoders leave distinct ``fingerprints,'' which can be used for source attribution \cite{2023arXiv230906780Z}, a critical capability for forensics \cite{ieee_forensics_2024}, copyright protection \cite{guo_usenix_2025}, and developing general defense strategies \cite{muller_interspeech_2022}.

\begin{figure*}[h]
    \centering
    \includegraphics[width=1\textwidth]{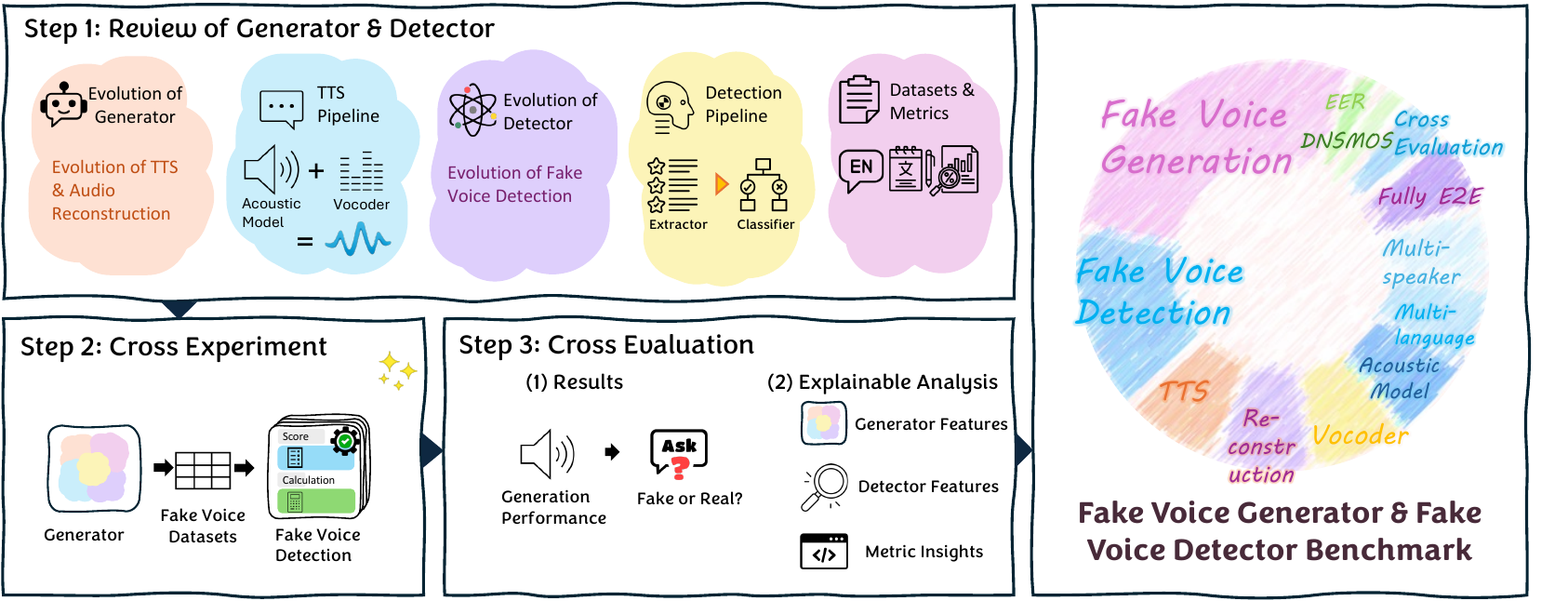}
    \caption{SoK pipeline for evaluating fake voice generation and fake voice detection systems which includes three parts: taxonomy of both fake voice generator and fake voice detector, cross experiments, and evaluation on both generator and detector. }
    \label{fig:pipeline}
\end{figure*}

We introduce an ecosystem-level benchmark that systematically evaluates the interplay between modern fake voice generation attack and detection defense. Our framework challenges 17 State-Of-The-Art (SOTA) fake voice generator, spanning TTS, TTS with voice conversion, and audio reconstruction, with 8 leading fake voice detectors using a novel one-to-one evaluation protocol. Unlike conventional assessments, our approach evaluates each generator-detector pair individually to expose unique acoustic artifacts, method-specific sensitivities, and previously hidden vulnerabilities. We proposed new unified metric for generator and detector to generally compare the performance of both in diverse dimensions, and we also clarify the underlying reasons for detector performance discrepancies, enabling the community to pinpoint techniques that threaten defenses. Figure \ref{fig:pipeline} outlines our evaluation pipeline.

\textbf{Our Contributions:}
\begin{enumerate}
    \item \textbf{Comprehensive Security Taxonomy:} We provide a detailed taxonomy of modern fake voice generation and detection systems, framed through a security lens to identify critical architectural vulnerabilities.

    \item \textbf{Large-Scale Empirical Assessment:} We conduct the most extensive cross-domain security evaluation to date, benchmarking 8 State-Of-The-Art detectors against 17 diverse fake voice generators. This one-to-one analysis reveals previously hidden, method-specific vulnerabilities that are obscured by traditional aggregated testing.
    \item \textbf{Unified Performance Metrics:} We propose a novel and unified scoring system to standardize the evaluation of both generator evasiveness and detector robustness, enabling fair comparisons across different technologies.

    \item \textbf{Actionable Recommendations:} Based on our extensive empirical findings, we offer concrete recommendations for developing the next generation of resilient detection systems and designing security-aware voice generators.
\end{enumerate}

\section{Fake Voice Generators}
\label{gen}

This section provides a taxonomy of fake voice generation systems, focusing on Text-to-Speech (TTS), the process of converting text into spoken voice. We particularly examine zero-shot voice conversion, a technique enabling voice generation from a short voice clip of a target speaker without specific prior training \cite{2023arXiv230102111W}.

\begin{figure}[h]
    \centering
    \includegraphics[width=\linewidth]{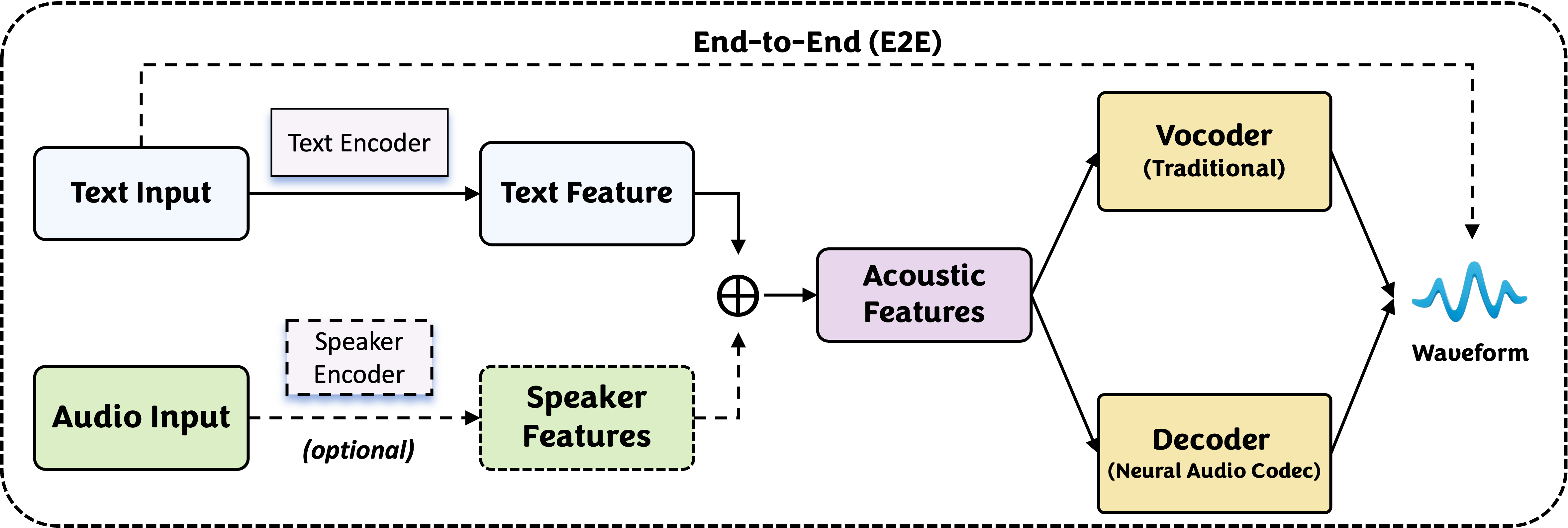} 
    \caption{The evolution of fake voice generation. Traditional neural TTS pipelines convert text to acoustic representations, then use vocoders to synthesize waveforms. Fully E2E models optimize all components at once. Neural codec models learn to encode and decode audio, replacing static compression algorithms. Audio inputs are often used for speaker conditioning in zero-shot TTS.}
    \label{fig:tts}
\end{figure}

The evolution of these systems is depicted in Figure \ref{fig:tts}. Fake voice generation has progressed from traditional two-stage pipelines to more integrated models. Traditional approaches use an acoustic model to convert text into an intermediate acoustic representation (like a spectrogram), which a vocoder then synthesizes into a waveform. These can be optimized as a single, fully end-to-end (E2E) model. A more recent paradigm involves neural audio codecs, which use a data-driven approach to compress and reconstruct voice. In this setup, a language model predicts a sequence of discrete voice tokens from text, which the codec's decoder then converts into the final waveform.

\subsection{Traditional Generators with Vocoders}
Traditional fake voice generators consist of two main components: an acoustic model and a vocoder.

\subsubsection{Acoustic Models}
Acoustic models generate acoustic representations from linguistic features \cite{2021arXiv210615561T}. Key architectures include:

\textbf{Autoregressive Models}: These models, pioneered by the Tacotron series \cite{2017arXiv170310135W}, generate spectrograms sequentially using encoder-attention-decoder architectures. While known for their naturalness, they face challenges with latency. Later models like DurIAN \cite{2019arXiv190901700Y} improved alignment stability by using duration predictors instead of attention mechanisms.

\textbf{Flow-Matching \& Diffusion-Based Models}: These models transform a simple distribution (e.g., Gaussian noise) into a complex data distribution. Flow-matching models learn a vector field to transport samples from a base to a target distribution, as seen in Matcha-TTS \cite{2023arXiv230903199M}, which enhances prosody control. Diffusion models, such as Grad-TTS \cite{2021arXiv210506337P}, gradually denoise samples from pure noise into coherent data, achieving high-fidelity synthesis through iterative refinement. Flow matching effectively unifies these deterministic (ODE-based) and stochastic (SDE-based) approaches \cite{lipman2023flow}.

\textbf{Transformer-Based Models}: The Transformer architecture \cite{2017arXiv170603762V} has become dominant in TTS due to its self-attention mechanism. Early examples like Transformer-TTS \cite{2018arXiv180908895L} adapted the architecture for voice. To address the latency of autoregressive methods, the non-autoregressive FastSpeech series \cite{2019arXiv190509263R, 2020arXiv200604558R} introduced parallel generation by predicting phoneme durations, pitch, and energy, which simplified training and improved control.

\subsubsection{Vocoders}
Vocoders synthesize raw voice waveforms from the acoustic features generated by the acoustic model.

\textbf{Autoregressive Vocoders}: Models like WaveNet \cite{2016arXiv160903499V} and WaveRNN \cite{2018arXiv180208435K} generate voice samples sequentially, where each sample is conditioned on previous ones. This approach captures temporal dependencies well, resulting in highly natural voice but at the cost of slow inference speed.

\textbf{Non-Autoregressive Vocoders}: To accelerate generation, several parallel vocoding methods have been developed:
\begin{itemize}
    \item \textbf{Flow-Based Vocoders}, such as WaveGlow \cite{2018arXiv181100002P}, use invertible transformations to map a simple distribution to a complex waveform, enabling fast, parallel synthesis.
    \item \textbf{Diffusion-Based Vocoders}, like DiffWave \cite{2020arXiv200909761K} and the more efficient RFWave \cite{2024arXiv240305010L}, generate voice by reversing a diffusion process that gradually adds noise. They offer high-quality synthesis with improving speeds.
    \item \textbf{GAN-Based Vocoders} employ adversarial training. Models like MelGAN \cite{2019arXiv191006711K} and the highly influential HiFi-GAN \cite{2020arXiv201005646K} use a generator and one or more discriminators to achieve a balance of high fidelity and computational efficiency widely used in modern TTS systems.
\end{itemize}

\subsection{Fully End-to-End Models}
\label{sec:e2e}
Fully end-to-end models generate waveforms directly from text, bypassing intermediate acoustic representations like spectrograms. This integrated approach simplifies training, avoids error propagation between stages, and can improve overall voice quality \cite{2021arXiv210615561T}. Key models such as VITS \cite{2021arXiv210606103K} use a conditional variational autoencoder with adversarial training to achieve high-fidelity E2E generation. Other notable contributions include Your-TTS \cite{2112.02418}, which pioneered multilingual zero-shot voice conversion, and NaturalSpeech \cite{2022arXiv220504421T}, which introduced novel components like a differentiable durator to enhance naturalness.

\subsection{Neural Audio Codec Language Models}
\label{sec:codeclm}
A leading paradigm in modern TTS involves neural audio codecs and language models. A neural audio codec (e.g., Encodec \cite{2022arXiv221013438D}) uses an encoder to compress waveforms into a compact sequence of discrete tokens and a decoder to reconstruct the voice from these tokens. A language model is then trained to predict these voice tokens from input text, often conditioned on a short voice for zero-shot voice conversion.

This approach was popularized by VALL-E \cite{2023arXiv230102111W}, which uses a Transformer-based language model to predict Encodec tokens. Subsequent models have advanced the generator performance:
\begin{itemize}
    \item \textbf{NaturalSpeech 2 \& 3 \cite{2023arXiv230409116S, DBLP:conf/icml/JuWS0XYLLST000024}:} Introduced a factorized codec design and continuous vector representations to achieve extremely high voice fidelity.
    \item \textbf{XTTS \cite{casanova2024xtts} \& CosyVoice \cite{2407.05407}:} Focused on massively multilingual support and fine-grained voice cloning capabilities.
    \item \textbf{VoiceCraft \cite{peng-etal-2024-voicecraft}:} Proposed a novel token rearrangement technique to improve synthesis quality and editing capabilities.
    \item \textbf{FireRedTTS \cite{2409.03283}:} Developed a streamable architecture with causal convolutions, enabling real-time applications.
\end{itemize}

The high fidelity achieved by these models, particularly their ability to convincingly clone voices from just a few seconds of voice, raises significant concerns about misuse and impersonation. This underscores the urgent need for robust fake voice detection and watermarking techniques to ensure the responsible deployment of these powerful technologies.
\section{Fake Voice Detectors}

To advance the field of fake voice detection, numerous studies \cite{2024arXiv240413914L,2024arXiv240915180P} have investigated fake voice detection techniques, which can be broadly classified into two categories: pipeline and end-to-end detectors. While pipeline approaches separate feature extraction and classification, End-to-End (E2E) models operate directly on raw voice waveforms. The following section introduces both frameworks in detail.

\subsection{Feature Extraction}

As the front-end and a crucial component of pipeline-based detection systems, feature extraction aims to derive discriminative representations by capturing artifacts indicative of fake voice. Depending on the feature generation method, existing approaches can be broadly divided into three categories: hand-crafted spectral features, Deep Learning (DL) features, and other analysis techniques.

\subsubsection{Hand-Crafted Spectral Features}
Hand-crafted spectral features leverage domain expertise to analyze a voice signal's properties in the frequency domain, providing a strong baseline for fake voice detection. These features, computed from the signal's power or magnitude spectrum, capture how its energy is distributed across different frequencies.

Prominent examples include Mel Frequency Cepstral Coefficients (MFCC) \cite{davis1980comparison}, Linear Frequency Cepstral Coefficients (LFCC) \cite{alegre2013one}, and two-dimensional spectrograms like the Mel-Spectrogram and Constant-Q Transform (CQT) Spectrogram \cite{fathan2022mel, cheng2019replay}. These representations are effective at highlighting discriminative artifacts indicative of fake speech. Subsequently, deep learning models such as Residual Networks (ResNet), Recurrent Neural Networks (RNNs), or U-Nets are often employed to process these features and generate deep embeddings \cite{khan2024frame, chen2023twice}.

A primary challenge, however, is robustness to real-world conditions. Empirical evidence from ASVspoof evaluations shows that performance can be significantly degraded by channel and device variability in physical access attacks \cite{asvspoof2017_evalplan,li2017_replay} and by codec artifacts in logical-access scenarios involving VoIP transmission \cite{asvspoof2021_evalplan}.
\subsubsection{Deep Learning Features}

Due to progress in Deep Learning (DL), neural network architectures are now employed to derive learnable embeddings that represent the essential traits of raw voice, complementing conventional hand-crafted features. While early approaches relied on traditional machine learning techniques and convolutional neural networks (CNNs), the field has evolved to incorporate more advanced architectures such as graph neural networks (GNNs), Transformers, and other modern deep learning models.

In recent years, self-supervised learning (SSL) frameworks have gained significant traction for feature extraction in fake voice detection, exhibiting remarkable advantages over conventional acoustic and supervised learning-based representations \cite{lv2022fake, 2022arXiv220212233T}. Xie et al. \cite{xie2021siamese} demonstrate the notable efficacy of Wav2Vec features when implemented within a Siamese network framework. Based on this study, Tak et al. \cite{2022arXiv220212233T} further explore the application of Wav2Vec-2.0 as a front-end feature extractor, combined with a spectro-temporal graph attention network (AASIST) as the back-end classification. Similarly, Guo et al. \cite{10447923} integrate the WavLM model with a Multi-Fusion Attentive classification to enhance fake voice detection capabilities. In contrast, Wang et al. \cite{wang2021investigating} employ embeddings from the pre-trained HuBERT as inputs.
Despite their accuracy, SSL embeddings and DL front-ends are vulnerable to adversarial perturbations and transfer-based black-box attacks that can evade state-of-the-art audio deepfake detectors, underscoring the need for attack-aware training and evaluation \cite{kawa2023_adv_defense,rabhi2024_adv_review,farooq2025_transferable}. This risk is especially salient in fraud/social-engineering use cases, where attackers can query or approximate deployed models \cite{ftc2023_voice_challenge}.

\subsection{Classification}

Traditional back-end classifiers for fake voice detection include conventional models such as Gaussian Mixture Models (GMMs) and Support Vector Machines (SVMs) \cite{reynolds1995robust, villalba2015spoofing}. With the advancements in deep learning, convolutional neural networks (CNNs) have gained prominence for their ability to extract localized and hierarchical patterns within voice signals. Wu et al. \cite{wu2020light} and Lavrentyeva et al. \cite{lavrentyeva2019stc} implemented the Light CNN (LCNN) architecture \cite{wu2018light} for synthetic voice identification, incorporating convolutional and max-pooling operations with Max-Feature-Map (MFM) activation to enhance discriminative feature selection.

The Residual Network (ResNet), a widely adopted CNN variant, addresses gradient vanishing issues in deep architectures through its skip-connection design. Both Alzantot et al. \cite{alzantot2019deep} and Tomilov et al. \cite{tomilov2021stc} employ ResNet as their classification component, achieving robust performance. Zhang et al. \cite{zhang2022sernet} introduced the Squeeze-and-Excitation Network (SENet) for fake voice detection, which enhances CNN representational capacity through channel-wise feature recalibration. ASSERT (ASqueeze-Excitation and Residual neTwork) \cite{lai2019assert}, a hybrid architecture combining SENet and ResNet modules, was recognized among the top-performing systems in the ASVspoof 2019 challenge.

Graph Neural Networks (GNNs) have emerged as effective approaches in fake voice detection by modeling frequency and temporal dimensions as graph nodes \cite{scarselli2008graph}. Graph Convolutional Networks (GCNs) segment spectrograms into grid-like patches from which CNN-extracted embeddings are derived \cite{chen2023graph}. Tak et al. \cite{tak2021graph} proposed a Graph Attention Network (GAT) incorporating attention mechanisms to improve detection accuracy. The AASIST framework \cite{tak2021graph} enhances GAT by integrating spectral and temporal sub-graphs through a heterogeneity-aware approach.

Zhang et al. \cite{zhang2024audio} introduce XLSR (Wav2Vec-2.0) with SLS Classifier, which employs a Sensitive Layer Selection module to extract discriminative features from transformer layers for fake voice detection. XLSR-Conformer \cite{rosello2023conformer} combines XLS-R with conformer architecture using classification tokens for variable-length utterances. Further refinement through XLSR-Conformer with TCM \cite{truong2024temporal} incorporates Temporal-Channel Modeling to capture dependencies in voice signals, enhancing fake voice detection.
In deployments such as telephony or VoIP, smart speakers, back-ends must generalize to unseen spoofing methods and environmental shifts; ASVspoof analyses consistently highlight generalization gaps from lab conditions to field conditions, motivating cross-corpus and cross-attack evaluation \cite{asvspoof2021_overview,kulkarni2024_unseen}. Classifiers should therefore be assessed not only on raw synthetic speech but also under compression, re-recording, and laundering processes that attackers use to obfuscate artifacts \cite{safe2025_challenge}.

\subsection{E2E Model}

End-to-end approaches, which operate directly on raw voice waveforms without requiring explicit feature extraction, have gained significant attention in voice-related tasks, such as fake voice detection. Among these, RawNet2 \cite{jung2020improved} is a neural network originally designed for speaker verification and identification. It processes raw voice signals using a sinc convolution layer for initial feature extraction, followed by multiple residual blocks to learn deeper representations. Tak et al. \cite{2020arXiv201101108T} were the first to adapt RawNet2 for anti-spoofing tasks. Building on this, Wang et al. \cite{wang2022audio} proposed a joint optimization strategy leveraging a weighted additive angular margin loss, further improving the performance of the RawNet2-based detection.

Another notable architecture, SincNet \cite{SincNet_DBLP:conf/slt/RavanelliB18}, is also designed to process raw voice directly, leveraging parameterized sinc functions to efficiently learn relevant frequency bands. Ge et al. \cite{ge2021raw} employed sinc-shaped filters and partial channel connections for efficiency. Zeinali et al. \cite{zeinali2019detecting} introduce a fusion framework that combines SincNet and VGG architectures to enhance fake voice detection performance. Additionally, Huang et al. \cite{10057965} propose DFSincNet, an end-to-end model that enhances SincNet by incorporating discriminative frequency information, with a particular focus on high-frequency components to improve spoofing detection accuracy. RawBMamba \cite{2406.06086} utilizes sinc and convolutional layers for short-range features and bidirectional mamba for long-range features, with a fusion module integrating embeddings to enhance voice representation. E2E detectors must be evaluated against adaptive adversaries and post-processing laundering (e.g., resampling, compression, playback–re-recording): recent benchmarks show notable performance drops when synthetic audio is intentionally processed to evade detection, and gradient-based attacks can reduce margins \cite{safe2025_challenge,wang2025_bim}.

\section{Cross-domain Evaluation}
To address the limitations of common fake voice detector evaluations, which typically rely on static and aggregated datasets, thereby obscuring the unique artifacts introduced by different fake voice generators, we propose a cross-domain, comprehensive assessment. This evaluation includes two experiments involving 17 types of fake voices, generated using distinct fake voice generator from three major categories: traditional pipeline models, neural codec language models, and end-to-end models, along with five existing datasets for comparison. To assess the robustness and generalizability of detection systems, we evaluate these voices using 8 different fake voice detectors, analyzing their performance on both individually generated voices and aggregated datasets. Since detector performance can vary significantly depending on the type of fake voice, with some detectors excelling in specific scenarios while failing in others, our approach provides a more nuanced understanding of their vulnerabilities. Additionally, we introduce two metric systems designed to quantify the overall effectiveness of both fake voice generators and detectors from multiple perspectives, enabling fair comparison and ranking across different systems.

\subsection{Experimental Design}

Experiments in this study were conducted using one NVIDIA RTX 4090 GPU (24 GB VRAM), an Intel i9-10900X CPU at 3.70 GHz, 256 GB of RAM, and running Ubuntu 22.04 LTS.

\subsubsection{Threat Model}
\begin{itemize}
    \item \textbf{Adversary's Goal:} To create fake voice that can successfully bypass automated fake voice detection systems, such as those used as countermeasures for ASV. The attacks include both zero-shot voice cloning and the generation of non-cloned but highly natural fake voice.
    \item \textbf{Adversary's Capabilities:} The adversary is assumed to have access to publicly available, State-Of-The-Art (SOTA) open-source fake voice generation models. They do not have white-box access to the target detectors but aim to create voice that is broadly evasive. This models a realistic scenario where sophisticated tools are readily accessible to a wide range of actors.
\end{itemize}

\subsubsection{Selection Criteria}
\textbf{Fake Voice Generators and Datasets:} Our selection of 15 generators was guided by three criteria: (1) \textbf{Architectural Diversity}, ensuring coverage of traditional pipelines, end-to-end models, and modern neural codec architectures as detailed in Section \ref{gen}; (2) \textbf{SOTA Performance}, including models recognized for producing high-fidelity output; and (3) \textbf{Public Accessibility}, using open-source models with pre-trained weights to ensure reproducibility. For existing datasets, we selected widely-used benchmarks (ASVspoof series, FoR) to ground our results in established evaluation paradigms. We included the CFAD dataset to test cross-lingual generalization. While other datasets like ADD \cite{2023arXiv230513774Y} exist, our evaluation focuses primarily on the English language. A detailed overview is in Table \ref{tab:select_models}.

\begin{table}[t]
  \centering
  \setlength{\tabcolsep}{0.8mm}
  \renewcommand{\arraystretch}{1.1}
  \caption{Details of selected fake voices from existing studies. NC-LM=Neural Codec Language Model. *: with Zero-Shot Voice Conversion Capability}
  \label{tab:select_models}
  \begin{tabular}{lll}
    \toprule
    \textbf{Category} & \textbf{Type} & \textbf{Models/Dataset} \\
    \midrule
    \multirow{7}{*}{TTS} & \multirow{2}{*}{Transformer-based}   & FastSpeech2 + HiFi-GAN \\
    & & YourTTS + HiFi-GAN * \\ 
    
    \cmidrule(lr){2-3}
    
    & Flow Matching           & Matcha-TTS + HiFi-GAN \\ 
    
    \cmidrule(lr){2-3}
    & \multirow{1}{*}{Diffusion-based}      & GRAD-TTS + HiFi-GAN \\ 
    
    \cmidrule(lr){2-3}
    
    & \multirow{2}{*}{NC-LM} & MaskGCT-TTS * \\
    & & FireRedTTS-1S * \\ 
    
    \midrule
    
    \multirow{10}{*}{Vocoder} & \multirow{3}{*}{GAN Structure} & HiFi-GAN \\
    & & MelGAN \\
    &  & ParallelWaveGAN \\ \cmidrule(lr){2-3}
    
    & \multirow{3}{*}{Arch Improved GAN} & Vocos \\
    & & Sifi-GAN \\
    & & BigVGAN \\ \cmidrule(lr){2-3}
    
    & \multirow{3}{*}{Diffusion-based} & Diffwave \\
    & & FastDiff \\
    & & RFWave \\
    
    \midrule
    
    \multirow{6}{*}{\begin{tabular}[c]{@{}l@{}}Existing\\Dataset\end{tabular}} & \multirow{4}{*}{English Dataset} & ASVSpoof-21LA \\
    & & ASVSpoof-21DF \\
    & & ASVSpoof-5 \\ 
    & & Fake or Real (FoR) \\ \cmidrule(lr){2-3}
    & Chinese Dataset                & CFAD \\
    \bottomrule
  \end{tabular}
\end{table}

\textbf{Fake Voice Detectors:} The 8 detectors were chosen based on: (1) \textbf{Strong Performance} on the public ASVspoof benchmark; (2) \textbf{Architectural Variety}, including both end-to-end models and those using powerful pre-trained front-ends like Wav2Vec-2.0; and (3) \textbf{Availability} of official open-source implementations. This ensures our evaluation reflects the current capabilities of leading detection paradigms. All selected detectors (Table \ref{tab:antispoofing_criteria}) were used with their official implementations, trained on the ASVspoof19-LA dataset.

\begin{table}[ht]
 \centering
 \setlength{\tabcolsep}{1mm}
 \renewcommand{\arraystretch}{1}
 \caption{Details of selected fake voice detectors from existing studies.}
 \label{tab:antispoofing_criteria}
 \begin{tabular}{lcc}
    \toprule
    \textbf{Category} & \textbf{Detection Approach} \\
    \midrule
    \multirow{4}{*}{End-to-End}
      & RawNet2 \cite{2020arXiv201101108T} \\
      & RawPCDart \cite{ge2021raw} \\
      & RawBMamba \cite{2406.06086} \\
      & AASIST \cite{2021arXiv211001200J} \\
    \midrule
    \multirow{4}{*}{Wav2Vec-2.0 Extraction}
      & XLSR-SLS \cite{zhang2024audio} \\
      & XLSR-Conformer \cite{rosello2023conformer} \\
      & XLSR-Conformer+TCM \cite{truong2024temporal} \\
      & XLSR-AASIST \cite{2022arXiv220212233T} \\
    \midrule
 \end{tabular}
\end{table}

\noindent \textbf{Fake Voice Generation Process.} LibriSpeech is a widely used open-source dataset designed for automatic speech recognition (ASR) and text-to-speech (TTS) research. It contains English voice derived from audiobooks, paired with aligned text transcriptions. In our experiments, we used the test-clean subset, which includes 8.56 hours of voice from 39 speakers, sampled at 24 kHz in a multi-speaker English corpus. For TTS systems, we used the transcriptions as input to generate fake voices. The original LibriSpeech voice served as the naturalistic reference. For TTS systems with zero-shot capabilities, able to mimic specific speaker characteristics, the generated fake voices closely resembled the original speaker. In contrast, TTS models without zero-shot functionality produced fake voices using a default or generic speaker profile. For audio reconstruction, we first converted the original waveforms into Mel-spectrograms and then back into waveforms using vocoders, treating the output as fake voice. We compared these reconstructed outputs with the original waveforms to evaluate the performance. All inference pipelines used in our experiments were sourced from publicly available pre-trained models on Hugging Face.

\subsection{Evaluating Fake Voice Generators}

In this section, we introduced the performance metrics used in our experiments and analyzed the correlations between them. We then proposed an overall performance score applicable to all types of generators, aimed at evaluating the effectiveness of both existing and future fake voices.

\subsubsection{Performance Metric for Generator}\label{subsec:generator-metrics}

The Mean Opinion Score (MOS) has been the standard for decades; yet, recent Text-to-Speech (TTS) systems often exceed the discriminative resolution of its five-point scale and suffer from listener bias, high variance, and costly test administration. \cite{le2024limits} Consequently, we adopted a set of objective metrics for TTS and vocoder evaluation that (i) correlate with perceptual quality, (ii) cover intelligibility, naturalness, and similarity, and (iii) computational overhead. Then, we computed an overall score based on those criteria. 

The performance of Text-to-Speech (TTS) systems, including zero-shot models, is evaluated using several metrics. \textbf{PSNR} \cite{2021arXiv211203099A} compares the maximum signal power to noise power, while \textbf{DNSMOS} \cite{9746108} uses deep learning to predict subjective audio perceptual quality. The \textbf{Real-time Factor (RTF)} measures efficiency, defined as the ratio of processing time to the audio's duration, where values under 1 signify faster-than-real-time performance. Model size and complexity are indicated by the total number of trainable \textbf{Parameters}. Intelligibility is assessed through \textbf{WER} (word error rate) and \textbf{CER} (character error rate) using Whisper \cite{10.5555/3618408.3619590}, and speaker similarity (\textbf{SIM}) is measured with a WavLM-based model \cite{9814838} between original and generated voice pairs. For vocoder models specifically, evaluation includes \textbf{PESQ} \cite{941023} to quantify perceptual differences, \textbf{SSIM} \cite{1284395} to assess frequency-domain similarity between spectrograms, and \textbf{LS-MSE} \cite{2021arXiv211203099A} to calculate the mean square error between the ground truth and generated log-mel spectrograms.

\subsubsection{Individual Metrics Analysis and Generator Overall Performance}
Each evaluation metric has its own strengths and limitations, making it suitable for specific applications. To create a more generalizable evaluation method, we introduce an overall performance score designed to assess the effectiveness of all types of voice generation systems, both current and future. Given the redundancy and complementarity among metrics used in Text-to-Speech (TTS) and vocoder models, we assign different weights to each metric and compute a final overall score to enable a more comprehensive performance analysis.

\noindent\textbf{Generator Overall Score Formulation.}
We compute an overall score as a weighted sum of normalized metrics:
\begin{align}
F_{\text{TTS}} &= \sum_{i} \alpha_i \cdot M_{i,\text{norm}}, \tag{1}\\[2pt]
F_{\text{Audio Reconstruction}} &= \sum_{j} \beta_j \cdot M_{j,\text{norm}}, \tag{2}
\end{align}
where $M_{\cdot,\text{norm}}\in(0,1]$ are per-metric values normalized as detailed in Appendix. 

For \textbf{TTS}, we assign highest mass to perceptual naturalness and speaker identity fidelity: DNSMOS is a no-reference proxy shown to correlate with human MOS for speech quality, while SIM captures speaker similarity, a standard proxy for identity preservation in zero-shot TTS with voice conversion. Intelligibility via WER is a long-standing ASR metric that tracks content accuracy. Fidelity metrics (PSNR) and structure-aware SSIM remain informative for reconstruction artifacts. Finally, deployment efficiency matters but should not dominate perceptual outcomes, so we give small weights to latency (RTF) and model footprint (LogParams).

For \textbf{Audio Reconstruction}, where speaker identity is typically not a target constraint, we shift mass to full-reference quality: DNSMOS complements reference-based measures; PESQ is the ITU-T standard for perceived speech quality; SSIM captures structural fidelity beyond pixelwise error; PSNR remains a widely used baseline; and efficiency terms, including RTF and LogParams, ensure practical deployability. This scheme prioritizes what end users perceive (naturalness or identity for TTS and perceptual fidelity for reconstruction) and includes efficiency as a secondary objective.

\definecolor{lightblue}{rgb}{0.88, 0.88, 1}   
\definecolor{lightred}{rgb}{1, 0.88, 0.88}    

\begin{table*}[ht]
  \centering
  \caption{Performance evaluation of various TTS generators. The best and worst performing values for each metric are highlighted in blue and red, respectively. WER and CER are shown in \%. Params = Parameters in Million, Overall = Overall score.}
  \label{tab:tts_generators_performance}
  \begin{tabular}{lccccccccc}
    \toprule
    Method & DNSMOS $\uparrow$ & WER $\downarrow$ & CER $\downarrow$ & PSNR $\uparrow$ & SIM $\uparrow$ & RTF $\downarrow$ & Params (M) & Overall $\uparrow$ \\
    \midrule
    FastSpeech2 + HiFi-GAN & 3.1140 & 18.41 & 8.52 & \cellcolor{lightred}20.4370 & \cellcolor{lightred}0.6650 & \cellcolor{lightred} 0.2403 & 71 & 0.291 \\
    GRADTTS + HiFi-GAN & \cellcolor{lightred}2.9018 & \cellcolor{lightred}21.41 & \cellcolor{lightred}9.05 & 22.5353 & 0.7147 & 0.0658 & 29 & 0.307 \\
    MatchaTTS + HiFi-GAN & \cellcolor{lightblue}3.3067 & 14.94 & 4.85 & 20.4433 & 0.7099 & 0.0871 & 21 & 0.586 \\
    YourTTS (zero-shot) & 3.2939 & 15.93 & 7.97 & \cellcolor{lightblue}22.8675 & 0.9189 & 0.2368 & 87 & 0.829 \\
    MaskGCT-TTS (zero-shot) & 3.2675 & 14.53 & 5.79 & 21.8595 & \cellcolor{lightblue}0.9567 & 0.3850 \cellcolor{lightred} & 2226 & 0.734 \\
    FireRedTTS-1S (zero-shot) & 3.2727 & 16.10 & 7.53 & 21.6621 & 0.9533 & \cellcolor{lightblue} 0.1768 & 1294 & 0.735 \\
    \midrule
    Ground Truth & 3.3414 & N/A & N/A & N/A & N/A & N/A & N/A & N/A \\
    \bottomrule
  \end{tabular}
\end{table*}

\definecolor{lightblue}{rgb}{0.88, 0.88, 1}   
\definecolor{lightred}{rgb}{1, 0.88, 0.88}    

\begin{table*}[ht]
  \centering
  \caption{Performance evaluation of various Vocoders (audio reconstruction). The best and worst performing values for each metric are highlighted in blue and red, respectively. Params = Parameters in Million, Overall = Overall score.}
  \label{tab:average_metrics_by_year_final_revised}
  \begin{tabular}{lccccccccc}
    \toprule
    Method & DNSMOS $\uparrow$ & PESQ $\uparrow$ & LS-MSE $\downarrow$ & PSNR $\uparrow$  & SSIM $\uparrow$  & RTF $\downarrow$ & Params (M) & Overall $\uparrow$ \\
    \midrule
    ParallelWaveGAN & 3.0895 & 2.8475\cellcolor{lightred} & 0.0012  & 25.2606 & 0.8478  & 0.0094 & 17  & 0.4541 \\
    MelGAN          & 3.0259 \cellcolor{lightred} & 2.9971 & 0.0007 & 26.7588 & 0.8831  & 0.0013 & 21  & 0.4716 \\
    HiFi-GAN        & 3.0372 & 3.1048 & 0.0006 \cellcolor{lightblue} & 26.7875 & 0.9418  & 0.0067 & 14  & 0.5607 \\
    DiffWave        & 3.1415 & 2.3865 & 0.0068 & 21.6858 & 0.7609  & 0.1152 & 3   & 0.2473 \\
    FastDiff        & 3.1211 & 3.2667 & 0.0093 \cellcolor{lightred} & 19.4235 \cellcolor{lightred}& 0.7550  \cellcolor{lightred} & 0.0439 & 15  & 0.3485 \\
    SiFi-GAN        & 3.0416 & 3.1653 & 0.0009 & 25.8332 & 0.9551  & 0.0081 & 11  & 0.5747 \\
    Vocos           & 3.2325 & 3.5996 & 0.0008 & 27.1873 & 0.9751  & 0.0705 & 14  & 0.7890 \\
    BigVGAN         & 3.2797\cellcolor{lightblue} & 4.0125 \cellcolor{lightblue} & 0.0007 & 28.3506 \cellcolor{lightblue} & 0.9812 \cellcolor{lightblue} & 0.0269  & 112 & 0.9081 \\
    RFWave          & 3.0553 & 3.8183 & 0.0089 & 19.7259 & 0.7598  & 0.0522 & 18  & 0.3495 \\
    \midrule
    Ground Truth    & 3.3414 & N/A    & N/A    & N/A     & N/A     & N/A & N/A & N/A \\
    \bottomrule
  \end{tabular}
\end{table*}
\noindent \textbf{Individual Metrics.} Table \ref{tab:tts_generators_performance} shows the performance of TTS and TTS with zero-shot voice conversion methods, while Table \ref{tab:average_metrics_by_year_final_revised} shows the detailed performance evaluation of audio reconstruction methods. For audio reconstruction, BigVGAN's strong performance across multiple perceptual (DNSMOS, PESQ) and signal-level metrics (PSNR) suggests a robust architecture capable of generating both realistic and accurate waveforms, potentially making its output more challenging for fake voice detection systems. Conversely, FastDiff's low PSNR and MelGAN's lower DNSMOS might reflect trade-offs that create detectable signatures for detection systems.

For TTS generators, Matcha-TTS achieving top DNSMOS suggests it excels in overall naturalness, raising the difficulty for detectors. For TTS with zero-shot voice conversion, MaskGCT-TTS and FireRedTTS-1S's top-level scoring in multiple metrics (e.g., SIM and PSNR) reflects their strong voice conversion ability and the perceptual fidelity that neural codec provides. YourTTS, as representative of end-to-end TTS with zero-shot conversion, achieves voice conversion with fewer parameters than neural codec language model TTS, supporting lightweight computational efficiency.

\noindent\textbf{Overall Performance Results.} Flow Matching approaches (MatchaTTS + HiFi-GAN) perform commendably with good DNSMOS scores. For ZSVC systems, YourTTS stands out with high overall score and competitive PSNR/SIM performance while maintaining moderate parameter count (87M). Neural Codec models like FireRedTTS-1S and MaskGCT-TTS deliver strong results but with substantially larger parameter counts (1294M and 2226M respectively). For vocoders, architecture-improved GANs such as BigVGAN and Vocos demonstrate leading performance, while diffusion-based generally exhibit lower overall scores despite fewer parameters.

\subsection{Cross-Domain Evaluation Performance Analysis}

We perform a comprehensive analysis based on Table~\ref{tab:combined_metrics_part1} and Table \ref{tab:combined_metrics_part2} with three foci. In Section \ref{sec:generalizationfake}, we discuss one generator performance across detector. In Section \ref{sec:generalizationdetect}, we discuss one detector performance across different generator. Then finally, we provide our overall score analysis in Section \ref{sec:overall} from the both ways to evaluate the overall detector performance.  
\definecolor{lightblue}{rgb}{0.88, 0.88, 1}  
\definecolor{lightred}{rgb}{1, 0.88, 1}    
\definecolor{lightgreen}{rgb}{0.88, 1, 0.88} 
\definecolor{lightgrey}{rgb}{0.85, 0.85, 0.85}


\begin{table*}
    \centering
    \setlength{\tabcolsep}{0.3mm}
    \caption{Performance Individual metrics (EER(\%), $C_{llr}$, minDCF) - Part 1. The best and worst performing values for each metric are highlighted in blue and red, respectively. All metrics are the lower the better.}
    \label{tab:combined_metrics_part1}

    \begin{tabular}{@{}l|ccc|ccc|ccc|ccc|ccc@{}}
        \toprule
        \multirow{2}{*}{\textbf{Detector}} & \multicolumn{3}{c|}{\textbf{Matcha‑TTS}} & \multicolumn{3}{c|}{\textbf{FastSpeech2}} & \multicolumn{3}{c|}{\textbf{GRAD‑TTS}} & \multicolumn{3}{c|}{\textbf{YourTTS}} & \multicolumn{3}{c@{}}{\textbf{MaskGCT‑TTS}} \\
        \cmidrule(lr){2-4} \cmidrule(lr){5-7} \cmidrule(lr){8-10} \cmidrule(lr){11-13} \cmidrule(lr){14-16}
        & EER & $C_{llr}$ & minDCF & EER & $C_{llr}$ & minDCF & EER & $C_{llr}$ & minDCF & EER & $C_{llr}$ & minDCF & EER & $C_{llr}$ & minDCF \\
        \midrule
        RawNet2           & 47.88 & \cellcolor{lightred}4.967 & 0.997 & 23.10 & 2.395 & 0.873 & 38.93 & \cellcolor{lightred}4.970 & 0.922 & 30.31 & \cellcolor{lightred}4.955 & 0.867 & 62.44 & 5.198 & 0.999 \\
        RawPCDart         & 48.36 & \cellcolor{lightblue}0.998 & \cellcolor{lightred}0.998 & 33.08 & \cellcolor{lightblue}0.991 & 0.915 & \cellcolor{lightred}49.97 & \cellcolor{lightblue}1.013 & \cellcolor{lightred}0.999 & \cellcolor{lightred}49.46 & \cellcolor{lightblue}0.998 & \cellcolor{lightred}0.988 & 54.81 & \cellcolor{lightblue}1.025 & \cellcolor{lightred}1.000 \\
        AASIST            & 41.70 & 1.639 & \cellcolor{lightred}0.998 & 14.24 & 2.393 & 0.843 & 37.69 & 1.614 & 0.939 & 42.67 & 1.673 & 0.982 & 56.90 & 5.134 & 0.998 \\
        RawBMamba         & \cellcolor{lightblue}24.29 & 2.330 & \cellcolor{lightblue}0.688 & \cellcolor{lightred}49.16 & \cellcolor{lightred}4.383 & \cellcolor{lightred}0.998 & 23.07 & 2.212 & 0.604 & 34.22 & 3.166 & 0.891 & \cellcolor{lightred}65.18 & \cellcolor{lightred}6.796 & 0.999 \\
        XLSR‑AASIST       & 35.85 & 2.259 & 0.868 & 41.22 & 2.242 & 0.922 & 3.42 & 2.397 & 0.073 & 28.57 & 1.662 & 0.587 & 45.19 & 2.573 & 0.977 \\
        XLSR‑Conformer    & \cellcolor{lightred}49.85 & 1.729 & 0.995 & 26.72 & 1.695 & 0.764 & \cellcolor{lightblue}2.31 & 1.672 & 0.066 & 15.49 & 1.682 & 0.419 & 41.85 & 1.732 & 0.906 \\
        XLSR‑SLS          & 48.36 & 3.095 & \cellcolor{lightred}0.998 & \cellcolor{lightblue}2.58 & 2.910 & \cellcolor{lightblue}0.070 & 44.34 & 3.065 & 0.996 & \cellcolor{lightblue}5.46 & 2.910 & \cellcolor{lightblue}0.149 & \cellcolor{lightblue}24.94 & 2.951 & \cellcolor{lightblue}0.626 \\
        XLSR‑Conf.+TCM    & 43.89 & 1.576 & 0.911 & 30.34 & 1.298 & 0.852 & 3.49 & 1.232 & 0.100 & 9.59 & 1.244 & 0.268 & 27.00 & 1.333 & 0.677 \\
        \midrule[\heavyrulewidth]
    \end{tabular}

    \begin{tabular}{@{}l|ccc|ccc|ccc|ccc|ccc@{}}
        \multirow{2}{*}{\textbf{Detector}} & \multicolumn{3}{c|}{\textbf{FireRedTTS‑1S}} & \multicolumn{3}{c|}{\textbf{RFWave}} & \multicolumn{3}{c|}{\textbf{SiFiGAN}} & \multicolumn{3}{c|}{\textbf{BigVGAN}} & \multicolumn{3}{c@{}}{\textbf{HiFi‑GAN}} \\
        \cmidrule(lr){2-4} \cmidrule(lr){5-7} \cmidrule(lr){8-10} \cmidrule(lr){11-13} \cmidrule(lr){14-16}
        & EER & $C_{llr}$ & minDCF & EER & $C_{llr}$ & minDCF & EER & $C_{llr}$ & minDCF & EER & $C_{llr}$ & minDCF & EER & $C_{llr}$ & minDCF \\
        \midrule
        RawNet2           & 60.25 & 5.095 & \cellcolor{lightred}1.000 & 32.57 & \cellcolor{lightred}2.110 & 0.797 & 25.22 & \cellcolor{lightred}2.057 & 0.613 & 43.69 & \cellcolor{lightred}2.218 & 0.997 & \cellcolor{lightred}46.07 & \cellcolor{lightred}2.239 & \cellcolor{lightred}0.998 \\
        RawPCDart         & 53.94 & \cellcolor{lightblue}1.327 & \cellcolor{lightred}1.000 & 44.69 & 0.991 & \cellcolor{lightred}1.000 & 44.40 & 0.964 & \cellcolor{lightred}0.998 & 44.44 & 0.974 & \cellcolor{lightred}0.999 & 44.31 & 0.984 & \cellcolor{lightred}0.998 \\
        AASIST            & 57.88 & 5.573 & \cellcolor{lightred}1.000 & 43.66 & 1.484 & 0.946 & \cellcolor{lightred}47.54 & 1.709 & 0.989 & 45.13 & 1.533 & 0.999 & 40.59 & 1.333 & 0.947 \\
        RawBMamba         & \cellcolor{lightred}71.25 & \cellcolor{lightred}6.805 & \cellcolor{lightred}1.000 & \cellcolor{lightred}49.33 & 2.034 & 0.952 & 44.82 & 1.626 & 0.925 & \cellcolor{lightred}48.60 & 2.034 & 0.977 & 41.15 & 1.492 & 0.923 \\
        XLSR‑AASIST       & 49.15 & 3.651 & 0.990 & 39.42 & 1.605 & 0.912 & 35.88 & 1.362 & 0.845 & 42.59 & 1.809 & 0.997 & 27.12 & 0.977 & 0.573 \\
        XLSR‑Conformer    & 46.57 & 2.036 & 0.964 & 37.65 & 1.012 & 0.892 & 30.42 & 0.951 & 0.769 & 32.24 & \cellcolor{lightblue}0.967 & 0.802 & 11.41 & 0.809 & 0.269 \\
        XLSR‑SLS          & 36.59 & 3.362 & 0.737 & 30.94 & 1.153 & 0.729 & \cellcolor{lightblue}14.37 & 0.991 & \cellcolor{lightblue}0.376 & \cellcolor{lightblue}25.92 & 1.111 & \cellcolor{lightblue}0.648 & \cellcolor{lightblue}5.16 & 0.925 & \cellcolor{lightblue}0.136 \\
        XLSR‑Conf.+TCM    & \cellcolor{lightblue}30.86 & 1.879 & \cellcolor{lightblue}0.712 & \cellcolor{lightblue}23.58 & \cellcolor{lightblue}0.764 & \cellcolor{lightblue}0.529 & 20.47 & \cellcolor{lightblue}0.695 & 0.529 & 30.36 & 0.974 & 0.719 & 6.44 & \cellcolor{lightblue}0.400 & 0.161 \\
        \bottomrule
    \end{tabular}
\end{table*}

\subsubsection{Performance Metrics for Fake Voice Detectors}
We adopt standard metrics from the ASVSpoof Challenge: Equal Error Rate (EER) for discrimination, and minimum normalized Detection Cost Function (minDCF) and Log Likelihood Ratio Cost ($C_{llr}$) for a combination of discrimination and score calibration. To synthesize these results, we propose a Detector Overall Score ($S_i$) that balances empirical performance against model complexity.

The score for detector $i$ is calculated as $S_i = \alpha P_i + (1-\alpha) \hat{C}_i$, where $P_i$ is the normalized empirical performance and $\hat{C}_i$ is the normalized model complexity penalty (Parameter). The empirical performance $P_i$ is a weighted average of the detector's performance against each generator $j$, calculated as $P_i = \sum_j w_j A_{ij}$, where $A_{ij}$ is the average normalized EER, minDCF, and $C_{llr}$.

\textbf{Justification of Parameters:} Here we provide a rationale for our parameter choices:
\begin{itemize}
    \item \textbf{Generator-Specific Weights ($w_j$):} The weights $w_j$ are designed to ensure a detector's final score is more significantly impacted by its performance against more evasive, high-quality generators. To achieve this systematically, the weight for each generator $j$ is derived from its \textbf{Challenge Score ($R_j$)}. We define this score based on a direct principle: the higher the quality of the synthetic voice, the more challenging it is to detect. Therefore, a generator's Challenge Score ($R_j$) is set to be directly proportional to its overall quality score ($F_{\text{TTS}}$ or $F_{\text{Audio Reconstruction}}$) that we calculated in the previous section. This method ensures that generators producing the most realistic audio, such as the high-scoring neural codec LMs, are assigned the greatest weight, thereby focusing the evaluation on the most potent security threats.
    \item \textbf{Trade-off Factor ($\alpha$):} We set $\alpha=0.8$ to heavily prioritize a detector's empirical performance ($P_i$) over its model size penalty ($\hat{C}_i$). In the security context of fake voice detection, effectiveness is paramount, while model efficiency is a secondary, albeit important, consideration for practical deployment. This choice reflects that priority.
\end{itemize}
This scoring methodology provides a transparent, albeit opinionated, framework for ranking detectors based on a holistic view of their robustness and efficiency.

\subsubsection{Generalization Analysis of Fake Voice Generators}\label{sec:generalizationfake}

This section analyzes how State-Of-The-Art detectors generalize across different families of fake voice generators. We discuss performance trends using the Equal Error Rate (EER, \%) metrics reported in Table~\ref{tab:combined_metrics_part1} and \ref{tab:combined_metrics_part2}.

\noindent\textbf{Aggregated Datasets.} While detectors perform well on standard benchmarks like ASVSpoof-21LA, with top systems achieving EERs below $2\%$ (e.g., $0.82\%$ for XLSR-AASIST), their performance degrades significantly under distribution shifts. EERs increase on compressed voice like ASVSpoof-21DF (e.g., XLSR-AASIST to $2.85\%$) and data with waveform degradations in FoR (XLSR-SLS at $4.07\%$). The challenge is most severe in cross-lingual settings, such as the Chinese CFAD dataset, where even robust models like XLSR-SLS see EERs rise to $13.06\%$, and weaker ones like AASIST approach chance-level performance ($49.11\%$). This highlights a critical weakness in out-of-distribution generalization.

\noindent\textbf{Audio Reconstruction: GANs and Diffusion Models.} Detectability of audio reconstruction models correlates strongly with architectural maturity.
\begin{itemize}
    \item \textbf{GAN-based Models:} Early GANs like MelGAN, HiFi-GAN, and ParallelWaveGAN are easily identified by exploiting artifacts like phase misalignment, with a top EER of $0.57\%$ (XLSR-SLS). However, advanced architectures that mitigate these flaws are far more challenging. Best-case EERs rise to $14.37\%$ for SiFiGAN, $17.49\%$ for Vocos, and $25.92\%$ for BigVGAN.
    \item \textbf{Diffusion-based Models:} A similar trend holds for diffusion vocoders. Early models like DiffWave and FastDiff leave residual high-frequency noise, making them detectable with low EERs (e.g., $1.00\%$ for XLSR-SLS against DiffWave). In contrast, newer variants like RFWave, which better suppress such artifacts, are significantly harder to detect, pushing the best EER to $23.58\%$.
\end{itemize}

\definecolor{lightblue}{rgb}{0.88, 0.88, 1}  
\definecolor{lightred}{rgb}{1, 0.88, 1}    
\definecolor{lightgreen}{rgb}{0.88, 1, 0.88} 
\definecolor{lightgrey}{rgb}{0.85, 0.85, 0.85}

\begin{table*}
    \centering
    \setlength{\tabcolsep}{0.3mm}
    \caption{Performance Individual metrics (EER(\%), $C_{llr}$, minDCF) - Part 2. The best and worst performing values for each metric are highlighted in blue and red, respectively. All metrics are the lower the better.}
    \label{tab:combined_metrics_part2}

    \begin{tabular}{@{}l|ccc|ccc|ccc|ccc|ccc@{}}
        \toprule
        \multirow{2}{*}{\textbf{Detector}} & \multicolumn{3}{c|}{\textbf{MelGAN}} & \multicolumn{3}{c|}{\textbf{DiffWave}} & \multicolumn{3}{c|}{\textbf{FastDiff}} & \multicolumn{3}{c|}{\textbf{Parallel‑WaveGAN}} & \multicolumn{3}{c@{}}{\textbf{Vocos}} \\
        \cmidrule(lr){2-4} \cmidrule(lr){5-7} \cmidrule(lr){8-10} \cmidrule(lr){11-13} \cmidrule(lr){14-16}
        & EER & $C_{llr}$ & minDCF & EER & $C_{llr}$ & minDCF & EER & $C_{llr}$ & minDCF & EER & $C_{llr}$ & minDCF & EER & $C_{llr}$ & minDCF \\
        \midrule
        RawNet2           & 39.21 & \cellcolor{lightred}2.174 & 0.933 & 41.92 & \cellcolor{lightred}2.206 & 0.985 & 26.96 & \cellcolor{lightred}2.061 & 0.695 & 40.52 & \cellcolor{lightred}2.191 & 0.936 & 43.71 & 2.219 & 0.951 \\
        RawPCDart         & \cellcolor{lightred}43.45 & 0.976 & \cellcolor{lightred}0.996 & 44.79 & 1.000 & \cellcolor{lightred}1.000 & \cellcolor{lightred}42.30 & 0.950 & \cellcolor{lightred}0.924 & \cellcolor{lightred}43.69 & 0.979 & \cellcolor{lightred}0.996 & \cellcolor{lightred}51.90 & 1.029 & \cellcolor{lightred}1.000 \\
        AASIST            & 22.10 & 0.682 & 0.618 & 34.90 & 1.133 & 0.917 & 36.85 & 1.215 & 0.821 & 29.86 & 0.939 & 0.806 & 44.77 & 1.522 & 0.953 \\
        RawBMamba         & 35.17 & 1.203 & 0.856 & \cellcolor{lightred}47.23 & 1.488 & 0.999 & 34.88 & 1.528 & 0.886 & 29.98 & 1.817 & 0.746 & 48.13 & \cellcolor{lightred}2.903 & 0.997 \\
        XLSR‑AASIST       & 7.64 & \cellcolor{lightblue}0.425 & 0.144 & 17.54 & 0.572 & 0.372 & 24.72 & 0.830 & 0.555 & 13.25 & \cellcolor{lightblue}0.484 & 0.257 & 40.64 & 1.698 & 0.915 \\
        XLSR‑Conformer    & 0.77 & 0.764 & 0.019 & 3.32 & 0.773 & 0.080 & 15.20 & 0.832 & 0.370 & 1.47 & 0.767 & 0.038 & 25.88 & 0.914 & 0.649 \\
        XLSR‑SLS          & \cellcolor{lightblue}0.57 & 0.914 & \cellcolor{lightblue}0.016 & \cellcolor{lightblue}1.00 & 0.914 & \cellcolor{lightblue}0.031 & \cellcolor{lightblue}4.51 & 0.924 & \cellcolor{lightblue}0.118 & \cellcolor{lightblue}0.90 & 0.914 & \cellcolor{lightblue}0.026 & \cellcolor{lightblue}17.49 & 1.018 & \cellcolor{lightblue}0.429 \\
        XLSR‑Conf.+TCM    & 0.74 & 0.352 & 0.021 & 1.49 & \cellcolor{lightblue}0.353 & 0.040 & 8.06 & \cellcolor{lightblue}0.422 & 0.198 & 10.13 & 1.015 & 0.286 & 20.87 & \cellcolor{lightblue}0.705 & 0.499 \\
        \midrule[\heavyrulewidth]
    \end{tabular}

    \begin{tabular}{@{}l|ccc|ccc|ccc|ccc|ccc@{}}
        \toprule
        \multirow{2}{*}{\textbf{Detector}} & \multicolumn{3}{c|}{\textbf{FoR}} & \multicolumn{3}{c|}{\textbf{CFAD}} & \multicolumn{3}{c|}{\textbf{ASVSpoof‑5}} & \multicolumn{3}{c|}{\textbf{ASVSpoof‑21LA}} & \multicolumn{3}{c@{}}{\textbf{ASVSpoof‑21DF}} \\
        \cmidrule(lr){2-4} \cmidrule(lr){5-7} \cmidrule(lr){8-10} \cmidrule(lr){11-13} \cmidrule(lr){14-16}
        & EER & $C_{llr}$ & minDCF & EER & $C_{llr}$ & minDCF & EER & $C_{llr}$ & minDCF & EER & $C_{llr}$ & minDCF & EER & $C_{llr}$ & minDCF \\
        \midrule
        RawNet2           & 17.77 & \cellcolor{lightred}4.309 & 0.507 & 43.20 & 1.699 & 0.934 & 36.04 & \cellcolor{lightred}4.094 & 0.827 & 9.50 & \cellcolor{lightred}1.274 & 0.258 & 22.38 & 3.154 & 0.625 \\
        RawPCDart         & \cellcolor{lightred}43.26 & 0.959 & \cellcolor{lightred}0.935 & 46.76 & 1.035 & 0.990 & \cellcolor{lightred}46.22 & 1.008 & \cellcolor{lightred}1.000 & \cellcolor{lightred}13.72 & 0.798 & \cellcolor{lightred}0.375 & 24.84 & 0.907 & \cellcolor{lightred}0.908 \\
        AASIST            & 12.32 & 1.084 & 0.312 & \cellcolor{lightred}49.11 & 1.356 & \cellcolor{lightred}0.997 & 29.12 & 4.001 & 0.711 & 11.46 & 0.851 & 0.329 & \cellcolor{lightred}26.75 & \cellcolor{lightred}4.440 & 0.616 \\
        RawBMamba         & 24.49 & 1.584 & 0.528 & 34.78 & 2.324 & 0.850 & 38.68 & 2.612 & 0.884 & 3.19 & 0.435 & 0.125 & 15.85 & 2.951 & 0.426 \\
        XLSR‑AASIST       & 6.83 & 1.313 & 0.160 & 16.03 & \cellcolor{lightblue}0.670 & 0.452 & 6.77 & 1.231 & 0.136 & \cellcolor{lightblue}0.82 & \cellcolor{lightblue}0.054 & \cellcolor{lightblue}0.023 & 2.85 & \cellcolor{lightblue}0.081 & 0.598 \\
        XLSR‑Conformer    & 4.40 & 1.199 & 0.120 & 15.69 & 1.158 & 0.427 & 1.98 & 1.039 & 0.049 & 1.88 & 0.458 & 0.054 & 3.09 & 0.543 & 0.082 \\
        XLSR‑SLS          & \cellcolor{lightblue}4.07 & 1.349 & \cellcolor{lightblue}0.116 & \cellcolor{lightblue}13.06 & \cellcolor{lightred}2.889 & \cellcolor{lightblue}0.337 & 1.73 & 1.247 & 0.048 & 2.87 & 1.199 & 0.078 & \cellcolor{lightblue}1.92 & 1.485 & \cellcolor{lightblue}0.050 \\
        XLSR‑Conf.+TCM    & 5.01 & \cellcolor{lightblue}0.603 & 0.141 & 14.69 & 0.953 & 0.388 & \cellcolor{lightblue}1.64 & \cellcolor{lightblue}0.473 & \cellcolor{lightblue}0.047 & 1.03 & 0.202 & 0.056 & 2.06 & 0.504 & 0.089 \\
        \bottomrule
    \end{tabular}
\end{table*}

\noindent\textbf{Flow Matching TTS.} Flow matching systems like Matcha-TTS pose a formidable challenge by modeling real data distributions with high fidelity. This technique effectively evades most detection mechanisms, resulting in poor detector performance. The lowest EER achieved is $24.29\%$ (RawBMamba), while many sophisticated detectors perform near chance level, with EERs approaching $50\%$.

\noindent\textbf{Neural Codec Language Model TTS}
This family, including MaskGCT-TTS and FireRedTTS-1S, represents the most difficult detection scenario. These models leverage a neural audio codec as a final polishing stage, explicitly optimizing for perceptual realism. Their advanced techniques—such as mask-predict paradigms, two-step decoding, and zero-shot voice conversion—produce voice with fewer artifacts and more natural prosody. Consequently, detector performance is severely compromised. The best EER against MaskGCT-TTS is $24.94\%$ (XLSR-SLS), and it rises to $30.86\%$ for the even more advanced FireRedTTS-1S (XLSR-Conf.+TCM). For these models, many detectors fail entirely, with EERs exceeding $60\%$.


\subsubsection{Generalization Analysis of Fake Voice Detectors}
\label{sec:generalizationdetect}

We evaluate detector performance across two main categories, analyzing their generalization trends, underlying causes, and inherent trade-offs.

\noindent\textbf{End-to-End Models}
End-to-end models generally exhibit limited generalization, characterized by high error rates and poor calibration when faced with diverse and unseen fake voice generators. Performance within this category is inconsistent; for instance, RawNet2 often struggles across most generators, while RawPCDart can excel against specific types. A common weakness is unstable calibration, with models like RawNet2 and RawBMamba exhibiting high $C_{llr}$ scores.

This brittleness arises because standard end-to-end training objectives (e.g., cross-entropy) do not explicitly enforce robustness to domain shifts. The convolutional front-ends used in models like RawNet2 \cite{2020arXiv201101108T} and RawBMamba \cite{2406.06086} learn features that are highly discriminative for in-domain data but fail to generalize to novel synthesis artifacts. Even AASIST \cite{2021arXiv211001200J}, despite its improved graph attention classifier, is ultimately constrained by its RawNet2-based encoder and shows more failures than those with more specialized architectures.

\noindent\textbf{Wav2Vec-2.0 Feature Extraction Models}
This category of models demonstrates highly context-dependent performance, achieving State-Of-The-Art results against certain generator families but failing significantly on others. For example, XLSR-AASIST excels on in-domain data like ASVSpoof-21LA ($0.82\%$ EER) but is less effective against modern TTS systems. This sensitivity to generator type is a defining characteristic: performance often collapses against advanced unseen systems (e.g., Matcha-TTS, FireRedTTS-1S) while remaining strong against older GAN or diffusion models. Among these models, XLSR-Conformer+TCM often demonstrates superior score calibration.

This behavior is rooted in strong architectural inductive biases that lead to specialization. The learned layer weighting in XLSR-SLS \cite{zhang2024audio} or the specific interplay of local and global features in XLSR-Conformer \cite{rosello2023conformer} become highly optimized for the artifact types seen during training. While this specialization is effective for known spectral or temporal discontinuities, it can render the models blind to the subtle, high-fidelity artifacts or unnatural prosody in advanced TTS outputs.

\begin{mdframed}[linewidth=0.5pt, backgroundcolor=gray!10]
\textbf{Takeaways.}
No single detector achieves universal robustness. Feature-engineered models excel against certain generators but fail on novel ones, while end-to-end models show broader but weaker generalization.
\end{mdframed}

\subsubsection{Detector Overall Performance}
\label{sec:overall}
The discrepancy between EER, $C_{llr}$, and minDCF highlights the importance of considering multiple metrics for a complete picture. Cases like XLSR-SLS (low EER, high $C_{llr}$ in CFAD evaluation) show good separation but unreliable scores, while RawPCDart (sometimes low $C_{llr}$, high EER/minDCF) shows potentially calibrated but highly indiscriminative separation. We also find generally audio reconstruction has better calibration performance than those of TTS synthesis possibly because of the mechanisms are directly reconstructing audio features which leads to better‐calibrated confident estimates. Shih et al. \cite{10446558} suggest detectors are securing onto specific, potentially unreliable cues.  When these cues are altered or masked \cite{10446558}, the detector's confidence/scoring mechanism likely breaks down, which would manifest as poor calibration. Those scenarios highlight limitations not captured by single metric alone, demonstrating the necessity of multiple metrics evaluations. 

\begin{table}[htbp]
  \centering
  \caption{Parameter count and Overall scores ($S_i$) for various fake voice detectors. Lower scores indicate better overall performance. \textbf{Bold} indicates the best value in each column.}
  \label{tab:detector_overall_scores_alpha_0_8_weighted_Aij}
  \begin{tabular}{lcc}
    \toprule
    Detector & Params (M) & Overall Score ($S_i$) $\downarrow$ \\
    \midrule
    RawPCDart & 24.50 & 0.7739 \\
    RawNet2 & 25.43 & 0.7030 \\
     RawBMamba & 0.72 & 0.5681 \\
    XLSR-AASIST & 317.84 & 0.5073 \\
      AASIST & 0.30 & 0.5053 \\
          XLSR-Conformer& 319.74 & 0.4239 \\
    XLSR-SLS & 340.79 & 0.3870 \\

    XLSR-Conf.+TCM & 319.77 & \textbf{0.3855} \\

    \bottomrule
  \end{tabular}
\end{table}

The overall detector scores ($S_i$) presented in Table~\ref{tab:detector_overall_scores_alpha_0_8_weighted_Aij} synthesize empirical accuracy ($P_i$) and model size penalty ($\hat{C}_i$), building upon the detailed generator challenges discussed in Section~\ref{sec:generalizationfake} and detector-specific traits from Section~\ref{sec:generalizationdetect}. With empirical performance heavily weighted ($\alpha=0.8$) and EER/minDCF prioritized within $P_i$, lower $S_i$ scores indicate superior overall standing. The XLSR-based models, exemplified by XLSR-Conformer+TCM ($S_i = 0.3855$) and XLSR-SLS ($S_i = 0.3870$), achieved the top ranks. Their excellent average $P_i$ values—reflecting strengths such as good calibration and effectiveness against many generator types—successfully offset their large model sizes (high $\hat{C}_i$), despite known challenges with the most advanced TTS systems (e.g., MaskGCT-TTS, FireRedTTS-1S) highlighted in earlier sections.

Conversely, very lightweight models like AASIST ($S_i = 0.5053$) and RawBMamba ($S_i = 0.5681$) attained competitive mid-tier scores. Their minimal $\hat{C}_i$ penalties compensated for what Section~\ref{sec:generalizationdetect} described as more modest or variable empirical performance (e.g., ``limited generalization'' or higher $C_{llr}$ for some end-to-end models). Detectors such as RawPCDart ($S_i = 0.7739$) and RawNet2 ($S_i = 0.7030$) ranked lower, as their less competitive $P_i$ scores were not sufficiently counterbalanced by their model complexities. These $S_i$ rankings thus provide a holistic but weight-dependent view, confirming that while the newest generator architectures pose significant detection hurdles (in Section~\ref{sec:generalizationfake}), detectors excelling in broad empirical accuracy can lead when model efficiency is a secondary but still influential consideration.

\section{Future Directions}
In this section, we outline concrete and forward-looking research avenues to navigate the escalating race between synthetic voice generation and detection, highlighting needs of both technical innovations and systemic strategies.

\begin{enumerate}
\item \textbf{Generator Design with Holistic Risk Assessment.}
Future research in voice generation will likely be dominated by powerful architectures like Neural Codec Language Models and end-to-end TTS. We advocate for a holistic evaluation framework that assesses not only model performance but also inherent security risks. Progress should be measured along four critical axes: (i) perceptual quality, (ii) signal and spectrum fidelity, (iii) computational efficiency, and (iv) biometric security risk (i.e., spoofing capability). Furthermore, responsible development must include a forensic analysis of potential artifacts to proactively inform the detection community.

\item \textbf{Advancing Detector Robustness.}
To counter sophisticated generators, detector research must prioritize generalization and reliability. We recommend focusing on three key areas:
\begin{itemize}
    \item \emph{Hybrid and Ensemble Designs.} Fuse powerful Self-Supervised Learning (SSL) front-ends (e.g., XLSR) with diverse, specialized back-ends. Such ensembles can capture complementary acoustic and spectral cues, leading to more robust decisions than single architecture.
    \item \emph{Principled Generalization Techniques.} Incorporate methods to improve out-of-distribution performance. Techniques such as domain generalization, adversarial training, few-shot learning, and disentanglement losses~\cite{2501.08238,2412.19279} are critical for mitigating the sharp performance drops observed on unseen generators.
     \item \emph{Reliable Score Calibration.} Develop detectors whose likelihood ratio scores remain well-calibrated under distribution shifts. To be practically useful in real-world applications, detectors must close the gap between achieving a low EER and a high decision cost ($C_{llr}$).
\end{itemize}

\end{enumerate}

\section{Related Work}

\noindent\textbf{Surveys and Benchmarks of Fake Voice Generation \& Detection.} Tan et al.~\cite{2021arXiv210615561T} and Zhang et al.~\cite{2023arXiv230313336Z} discuss the rapid progress in neural TTS and diffusion models, respectively.  For fake voice detection, Khan et al.~\cite{2210.00417} and Li et al.~\cite{2024arXiv240413914L} offer taxonomies and pipeline analyses up to early 2024.  Literature such as VocBench~\cite{2021arXiv211203099A} and VoiceWukong~\cite{2024arXiv240906348Y} provide standardized benchmarking, while CodecFake+ broadens the corpus to unseen neural codec attacks and shows that diverse codec training improves robustness~\cite{2501.08238}.

\noindent\textbf{Generalization and Cross Evaluation.} Performance often collapses when detectors face unseen generators or corpora~\cite{2203.16263,10446558,2501.13887,10448016}. Doan et al.~\cite{10.1145/3658644.3690311} identify highly transferable artifacts from large E2E-GAN generators, and Mishra et al.~\cite{2502.04049} use Shapley-based explanations to pinpoint shortcut features.  Wu et al.’s CodecFake~\cite{wu} confirms that retraining on neural codec enhances cross-dataset resilience.

\section{Conclusion}
Our systematic evaluation shows that State-Of-The-Art fake voice generators consistently bypass today’s leading fake voice detectors, highlighting a fundamental trade-off between fidelity and security. At the same time, no single detection method proved robust against every fake voice generation techniques we tested, underlining serious gaps in current fake voice detection practices. To secure the future voice ecosystem, fake voice generator designs could embed forensic-aware constraints to be more secured while detector research must prioritize adaptability and cross-method generalization to stay ahead of a rapidly diversifying threat landscape.

\bibliographystyle{IEEEtran}
\bibliography{reference}

\end{document}